# Truly Sub-Nyquist Generalized Eigenvalue Method with High-Resolution

Baoguo Liu, Huiguang Zhang, Wei Feng, Zongyao Liu, Zhen Zhang, Yanxu Liu

*Abstract*—**The realization of spectral super-resolution sensing is of great significance in several applications, including radar, remote sensing, and wireless communication. However, in conventional sub-Nyquist sampling techniques, spectrum leakage and the picket-fence effect pose considerable challenges for the accurate extraction of super-resolution signal components, while the hardware implementation of random sampling is another crucial factor that hinders the widespread application of conventional sub-Nyquist sampling techniques. This study addresses these problems, by proposing a generalized eigenvalue method that exploits the incoherence between signal components and the linearity-preserving property of filtered operations. The generalized eigenvalue method enables the accurate extraction of signal component parameters with super-resolution under sub-Nyquist sampling conditions. Furthermore, the proposed method is based on uniform sub-Nyquist sampling, which is truly sub-Nyquist sampling and effectively reduces the difficulty of hardware implementation. Moreover, unlike traditional compressed sensing techniques, the proposed method does not rely on the discrete Fourier transform framework, thereby effectively eliminating spectral leakage and the picket-fence effect. Moreover, it greatly reduces the adverse effects of random sampling on signal reconstruction and hardware implementation.**

## I. INTRODUCTION

### *A. Overview of Sub-Nyquist Sampling Techniques*

Ultra-wideband and high-throughput signals are becoming increasingly common in fields such as wireless communication[1], radar [2], and computational imaging [3]. This trend poses unprecedented challenges in signal sampling, transmission, and processing. However, both artificial and natural signals exhibit energy distributions that frequently adhere to a power law pattern[4],[5]. This characteristic implies that many signals can be regarded as sparse, a trait that is particularly evident in wideband signals. Consequently, it becomes feasible to implement sub-Nyquist sampling techniques by utilizing prior knowledge.

Manuscript received **** **, 2024; revised ******** **, 2024 and ******* **, 2024; accepted ***** **, ****. Date of publication ******** **, ****; date of current version ***** **, 2024.This work was supported by National Natural Science Foundation of China (U1604254, 11172092, 10872064),.

BaoGuo Liu, Huiguang Zhang, Wei Feng and Zongyao Liu are with School of Mechanical and Electrical Engineering, Henan University of Technology, Zhengzhou, China (e-mail: liubaoguo1979@sina.com).

Zhen Zhang, Yanxu Liu is with the School of Mechanical Engineering, Zhengzhou University of Technology, Zhengzhou, China.

Currently, several primary sub-Nyquist sampling techniques are used including compressed spectrum sensing (CSS), modulated wideband converter (MWC), and sparse Fourier transform (SFT), and compressed covariance sensing (CCS).

CSS is a technique that combines compressive sensing (CS)[6],[7],[8]and Fast Fourier Transform (FFT) to efficiently process signals by converting non-sparse time-domain signals into sparse frequency-domain signals. The method is particularly useful in applications like cognitive radio, where it helps in detecting spectrum holes with reduced sampling rates. However, the accuracy of CSS can be affected by the basis mismatch problem, primarily due to spectral leakage in FFT, which can lead to inaccuracies in signal reconstruction.

Eldar *et al.* proposed the MWC method for sparse multiband signals. The MWC ensures signal recoverability by separating different band signals and performing random modulations [9]. This enables the MWC method to sample and reconstruct signals at frequencies much lower than the Nyquist rate, even when the carrier frequency is unknown [9][10]. These MWC methods have recently been regarded as crucial breakthroughs in sub-Nyquist sampling theory[11][12]. However, with growing sub-bands and bandwidths, the system complexity and power consumption of MWC may increase accordingly, posing challenges in practical application.

The sparse Fourier transform (SFT) proposed by Hassanieh et al. [13] provides an efficient sub-Nyquist method to process large-scale signal sets, which is suitable for the fast spectral analysis of large datasets, such as those at the terabyte level. SFT effectively reduces the number of samples required for extracting the main components by employing the barrel filter technique[14][15]. However, the lack of a priori knowledge and the selection of bucket coefficients and filter parameters pose formidable challenges, primarily because inappropriate parameters can degrade the extraction accuracy.

Regarding computational cost, Romero *et al.* [16]introduced CCS to address the processing requirements involved in reconstructing wideband analog signals. CCS involves developing a covariance matrix of the wideband signal from compressed samples, rather than directly reconstructing the signal itself [17]. The method efficiently minimizes the computational burden, which simplifies the process. However, while CCS simplifies computation, it imposes certain constraints. Unlike compressed sensing techniques that can precisely reconstruct the original signal, CCS may not achieve full signal reconstruction. Nevertheless, traditional sub-Nyquist sampling techniques are predominantly based on random sampling coupled with the discrete Fourier transform. The presence of inherent problems such as spectral leakage within the discrete Fourier transform, coupled with the influence of noise and other external factors, often leads to mismatches in the basis, which adversely affects the effectiveness of the reconstruction techniques[18],[19]. This deficiency is pronounced in high-precision contexts where the inherent limitations of these techniques are clearly expressed.

In terms of the hardware implementation of sub-Nyquist sampling technology, a considerable amount of related work has been conducted, resulting in many significant achievements. Eldar *et al.* made crucial advancements in hardware implementation. They developed a hardware system that can perform sub-Nyquist sampling and reconstruction of wideband signals via a MWC [11][20]. The study confirmed the practicality of the MWC technique, enabling the sampling and processing of signals within a subspace union. Pfetsch *et al.* presented a potential hardware implementation of an analog-to-information converter (AIC)[21]. The AIC performs sub-Nyquist sampling and reconstructs analog signals by modulating the incoming analog signal using pseudo-random sequences [22][23] and then extracting the wideband signal by compressing the measurement matrix. The switching frequency of the PR component within the AIC must comply with Nyquist's theorem.

Sub-Nyquist sampling techniques, which predominantly utilize random sampling, require the development of specialized PR hardware. This necessity increases the complexity of hardware development and limits the broad application of these technologies. To address this challenge, Li Xiao introduced a Nyquist sampling method based on multi-level uniform sampling, leveraging the Chinese Remainder Theorem and the FFT method, thus effectively mitigating the identified problem [23]. However, the method does not completely circumvent the effects of FFT spectrum leakage. In addition, subspace methods, including ESPRIT[24],[25],[26], MUSIC [27],[28], and those based on the generalized eigenvalue theory such as the Pencil-of-Functions method [29],[30] and the Matrix Pencil method [31],[32], have demonstrated superior performance in extracting high-precision frequencies. These methods have found extensive use in areas like direction of arrival (DOA) estimation and linear spectrum estimation. However, the MUSIC method necessitates a spectral peak search across the entire frequency band and is primarily appropriate for scenarios that comply with the Nyquist sampling theorem. Furthermore, methods grounded in the generalized eigenvalue theory are incompatible with Nyquist sampling owing to the periodic nature of discrete complex exponentials. These methods also fall short in accurately extracting amplitude and phase, an ability that is critical in specific applications such as power harmonic detection and inverse synthetic aperture radar (ISAR) migration alignment.

Note that the principal objective of this study is not to provide an exhaustive review of the field of sub-Nyquist sampling techniques. Our aim is to present a concise overview of the significant advances and central representative works in sub-Nyquist sampling theory, thereby enabling readers to quickly develop the requisite knowledge framework to comprehend the methodologies proposed in this study. Consequently, the literature review presented in this section may not encompass all aspects of sub-Nyquist sampling techniques.

*B. Current Challenges in SubNyquist methods*

As previously stated, although remarkable progress has been made in sub-Nyquist sampling techniques, the impact of spectral leakage and the picket-fence effect on the accuracy of spectral sensing warrants further study. Additionally, the subsampling algorithm under the traditional uniform sampling condition, that is, truly sub-Nyquist sampling, rather than an algorithm based on random sampling, is also under-researched.

1) **Non-Sparsification of Signals Caused by Discrete Fourier Transform (DFT) Spectral Leakage and Picket-Fence Effect:**
The conventional sub-Nyquist sampling techniques generally leverages Fourier transform matrices and random matrix theory to derive an measurement matrix. However, the spectral leakage and picket-fence effect of the DFT may cause the signal to lose its original sparsity. Current research approaches this problem in two methods: by assuming that the signal is standard sparse, or by enforcing sparsity through hard thresholding. This undoubtedly affects the super-resolution extraction of the frequency, amplitude, and phase of the reconstructed signal[18][23].
2) **Algorithms that Utilize Random Sampling Versus Truly Sub-Nyquist Sampling:**
Currently, random sampling methods are extensively employed in compressed sensing to ensure the restricted isometry property(RIP) of the sensing matrix. However, despite operating below the required sampling frequency of the Nyquist sampling theorem, random sampling poses two severe challenges. First, considering the computational complexity of reconstructing the signal after random sampling, there remains a certain probability of failure, although it is relatively low. These limitations render compressed spectrum sensing primarily suitable for application scenarios with high sampling costs, relatively sufficient computational resources, and low real-time and resolution requirements[24]. Second, the hardware implementation of random sampling presents a severe challenge, particularly to analog devices. This exists because the majority of existing commercial devices are designed for uniform sampling, rather than random sampling. Furthermore, there is a paucity of studies on algorithms based on truly sub-Nyquist sampling with uniform sampling[33],[34],[35].

Solving the aforementioned problems is of immense practical importance for the viability of super-resolution extraction and hardware implementation. Currently, there is a critical need to investigate super-resolution algorithms that can overcome the restrictions of FFT in truly sub-Nyquist sampling.

### *C. Main Achievements and Innovations*

In conventional sub-Nyquist sampling techniques, spectral leakage and picket-fence effect are key problems that affect super-resolution extraction, while difficulties in hardware implementation caused by random sampling have severely limited the application of sub-Nyquist sampling techniques.

To address these problems, inspired by the work of Sarkar et al. on matrix pencils, we propose a novel generalized eigenvalue method called the Sub-Nyquist generalized eigenvalue method (SNGEM). SNGEM leverages the non-coherence and linear mapping relationship between signal components and their filtered signals, along with a synchronized aliasing mechanism. Compared to the matrix pencil method and other sub-Nyquist methods. SNGEM enables the super-resolution extraction of component frequencies, amplitudes, and phases of multi-frequency complex signals at truly sub-Nyquist sampling rates or even at arbitrary sampling rates. The key contributions of this paper are as follows:

1) **Innovation in Theoretical Framework:**

   SNGEM can be applied beyond the Fourier transform framework, which enhances precision in parameter extraction without being influenced by spectral leakage or the picket-fence effect. Moreover, SNGEM employs uniform sampling, which reduces the impact of random sampling noise and theoretically enables the extraction of signal components at frequencies below the Nyquist limit. Super-resolution can be achieved by sampling parameters at sub-Nyquist rates.
2) **Optimization of Sampling Length:**
   Unlike other sub-Nyquist sampling techniques, the sample length in the SNGEM method is determined solely by the inherent sparsity of the signal and is independent of the original length of the signal. This greatly reduces the size of the sampled data, enhancing the effectiveness of the algorithm and reducing system requirements.
3) **Precise Extraction of Parameters of Linear Frequency-Modulation Signals:**
   This paper extends SNGEM to the multi-parameter eigenvalue problem (MEP) for the first time and successfully applies it to the extraction of parameters of linear frequency-modulation signals. This achievement enables blind and accurate parameter extraction at sub-Nyquist sampling rates.
4) **Hardware Implementation Simplification:**
   SNGEM employs a low-speed, uniform sampling approach, which simplifies its practical implementation using commonly utilized circuit components. Consequently, the complexity and cost of the required hardware are reduced.

The SNGEM approach improves the precision of sub-Nyquist sampling techniques considerably and simplifies hardware implementation. These findings offer innovative insights and solutions for signal processing, thereby improving sub-Nyquist sampling techniques to leverage their distinct advantages in real-world applications.

### *D. Organization*

The second section focuses on conducting a separate investigation and proving the generalized eigenvalue extraction theory for multi-frequency and linear frequency-modulation signals. The third section presents numerical test results to compare the performance of the SNGEM and FFT, evaluating the super-resolution spectrum sensing capability of the SNGEM under truly sub-Nyquist sampling with an ultra-wideband (UWB) signal. The fourth section elucidates the fundamental principles of sub-Nyquist sampling implementation of this algorithm and explores its potential for hardware implementation. The findings exemplify the innovative applications of SNGEM in signal processing and establish a theoretical foundation for future hardware implementation.

## II. Parameter extraction theory for Multi-Frequency Complex Signals and linear frequency-modulation based on generalized eigenvalue method

### *A. Multi-frequency Signal Parameter Extraction Theory*

1) **Theory of Component Frequency Extraction For Multi-Frequency Complex Signals**

**Definition 1.** A multi-frequency complex signal $x(t)$ is defined as follows:

$$x(t) = \sum_{i=1}^{m} a_i \, e^{j(2\pi f_i t + \varphi_{i0})}. \tag{1}$$

where $f_i, a_i, and\ \varphi_{i0}$ denote the frequency, amplitude, and initial phase of the *i*-th component, and $m$ indicates the number of components.

Once the signal $x(t)$ is processed by a filter with a certain frequency response characteristic (e.g. a Gaussian filter or a differential filter), the resulting signal is $\psi(t)$. The selected filter should be able to reliably and effectively capture the frequency response curve of the signal within the band of interest. Concurrently, the amplitude-frequency response coefficient should only correspond to a finite number of frequency points and, if possible, have a one-to-one mapping.

The Hankel matrix discretized and constructed for $x(t)$ is $X$,and that discretized and constructed for $\psi(t)$ is $\Psi$, where

$$X = \begin{pmatrix} x_1 & x_2 & \cdots & x_n \\ x_2 & x_3 & \cdots & x_{n+1} \\ \vdots & \vdots & \ddots & \vdots \\ x_n & x_{n+1} & \cdots & x_{2n-1} \end{pmatrix}, \Psi = \begin{pmatrix} \psi_1 & \psi_2 & \cdots & \psi_n \\ \psi_2 & \psi_3 & \cdots & \psi_{n+1} \\ \vdots & \vdots & \ddots & \vdots \\ \psi_n & \psi_{n+1} & \cdots & \psi_{2n-1} \end{pmatrix}. \tag{2}$$

Construct the generalized eigenvalue equation of $X$ to $F_x$:

$$\Psi u = \lambda X u, \tag{3}$$

where $\lambda,\ u$ represent the generalized eigenvalues and generalized eigenvectors of $\Psi$ to $X$.

**Proposition 1.** When $n \geq m$, the generalized eigenvalue equation, expressed as (3), for the multi-frequency complex signal yields $m$ non-zero complex generalized eigenvalues $\lambda_i$, given by the following equations:

$$\lambda_i = A_i(f_i) \exp\left(j\Phi_{i0}(f_i)\right). \tag{4}$$

Where $A_i(f_i)$ represents the amplitude frequency response function, $\Phi_{i0}(f_i)$ represents the phase frequency response function. Here we only give the key steps, please refer to the appendix for detailed proof.

Performing a Vandermonde decomposition of the Hankel matrix $\Psi$ with discrete components $\psi(t)$: $\Psi$ can be expressed as follows:

$$\Psi v - \lambda X u = S_c^T \begin{pmatrix} (\lambda-\beta_1)s_1 & & & \\ & (\lambda-\beta_2)s_2 & & \\ & & \ddots & \\ & & & (\lambda-\beta_m)s_m \end{pmatrix} S_c u, \qquad (5)$$

where $S_c$ represents the Vandermonde matrix of the Hankel matrix $X$ formed by the discrete multiple-frequency complex signal. For the detailed structure of Sc and the proof, please refer to Appendix A

Equation (5) shows that the rank of $\Psi - \lambda X$ decreases from $m$ to $m$–1 only when $\lambda = \beta_i$. This implies the existence of a vector $v$ corresponding to $\beta_i$, which satisfies the condition for a non-zero solution to $(\Psi - \lambda X)u = 0$. According to the definition of generalized eigenvalues, $\beta_i$ represents the generalized eigenvalue of (3). Therefore, the proof of Proposition 1 is complete.

Then for $\beta_i$, given a specific filter, since its amplitude-frequency characteristic is always known, we can extract the corresponding frequency $f_i$ by performing the inverse transformation of the amplitude $A(f_i)$ of the generalized eigenvalue $\beta_i$. In practical applications, the signal order information is often incomplete, resulting in a discrepancy between the number of sampling points and the signal order. Furthermore, it is difficult to circumvent the influence of noise on generalized eigenvalues. To address these issues, we use the singular value decomposition (SVD) method identify the relevant generalized eigenvalues[32].

Although the aforementioned proof is based on multi-frequency complex signals, it is straightforward to extend the application of this method to real multi-frequency signals via Euler's formula. A more intuitive understanding of SNGEM can be gained by Fig. 1. Here we use a first-order Butterworth filter. See III.A for its feasibility analysis.

The process flow of SNGEM is illustrated in Fig. 1 and summarized below:

First, the original signal $x(t)$ is differentiated by an analog filtered device to obtain the filtered signal $\psi(t)$. The original signal $x(t)$ and the filtered signal $\psi(t)$ are synchronously sampled at the same sampling rate (either satisfying the Nyquist sampling rate or not satisfying it). $\Psi$ are constructed. Subsequently, the generalized eigenvalue with Vandermonde equations in (3) are constructed according to $X$ and $\Psi$. Finally, the parameters are obtained by solving (3).

2) **Theory of Component Magnitude And Phase Extraction For Multi-Frequency Complex Signals**

**Definition 2.** A multi-frequency complex signal $g(t)$ is defined as follows:

$$g(t) = \sum_{i=1}^{m} e^{j2\pi f_i t} \qquad (6)$$

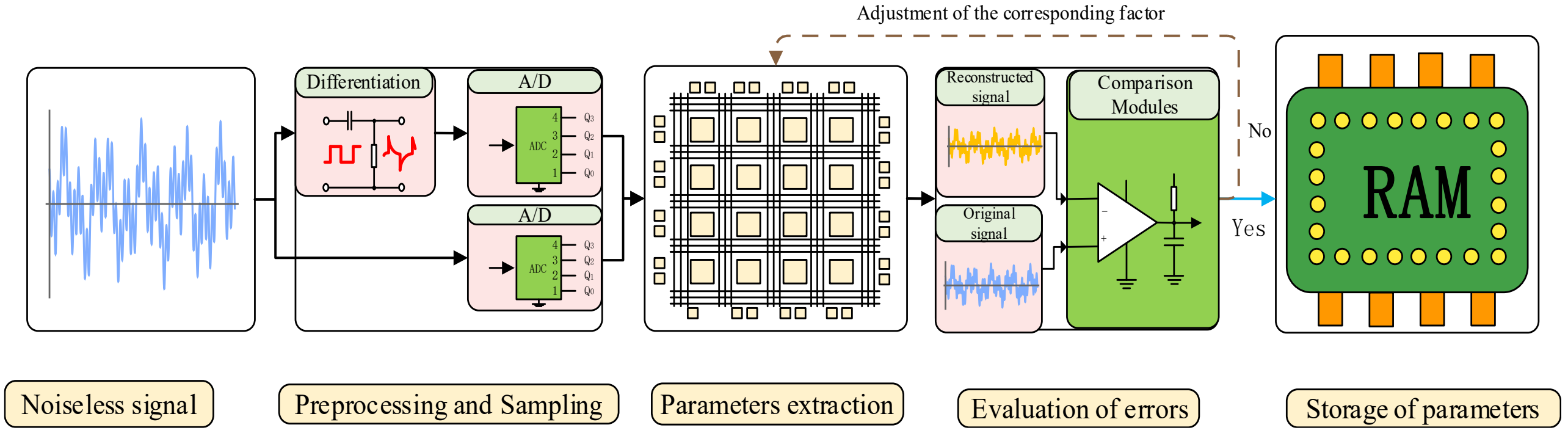

**Fig. 1** Flowchart of the hardware implementation of the generalized eigenvalue method.

The Hankel matrix $G$ is formed from the discretized $g(t)$.

$$G = \begin{pmatrix} g_1 & g_2 & \cdots & g_n \\ g_2 & g_3 & \cdots & g_{n+1} \\ \vdots & \vdots & \ddots & \vdots \\ g_n & g_{n+1} & \cdots & g_{2n-1} \end{pmatrix}, \qquad (7)$$

The Hankel matrix constructed from the discrete components of $g(t)$ is denoted as $G$.

**Corollary 1.** Once the frequencies of the components are determined via Proposition 1, the amplitude and phase of the multi-frequency signal are obtained using (8) [36][37].

$$a_i e^{j\varphi_{i0}} = \frac{u_i^H X u_i}{u_i^H G_i u_i} \qquad (8)$$

**Proof.** For simplicity, (7) can be expressed as follows:

$$g(t) = \sum_{i=1}^{m} e^{j\omega_i t} \qquad (9)$$

Thus, from (7) and (9), the constructed Hankel matrix $G$ with a discrete $g(t)$ is obtained as follows:

$$G = \begin{pmatrix} \sum_{i=1}^{m} s_i & \sum_{i=1}^{m} s_i^2 & \cdots & \sum_{i=1}^{m} s_i^n \\ \sum_{i=1}^{m} s_i^2 & \sum_{i=1}^{m} s_i^3 & \cdots & \sum_{i=1}^{m} s_i^{n+1} \\ \vdots & \vdots & \ddots & \vdots \\ \sum_{i=1}^{m} s_i^n & \sum_{i=1}^{m} s_i^{n+1} & \cdots & \sum_{i=1}^{m} s_i^{2n-1} \end{pmatrix} \qquad (10)$$

For all components $G_i$ of $G$, it holds that:

$$G_i = \begin{pmatrix} s_i & s_i^2 & \cdots & s_i^n \\ s_i^2 & s_i^3 & \cdots & s_i^{n+1} \\ \vdots & \vdots & \ddots & \vdots \\ s_i^n & s_i^{n+1} & \cdots & s_i^{2n-1} \end{pmatrix} \qquad (11)$$

Thus, from (3),(9), and (11), it follows that

$$u_i^H X_i u_i = a_i e^{j\varphi_{i0}} u_i^H G_i u_i \tag{12}$$

For the $i$-th generalized eigenvector $u_i$ , $u_i^H X_i u_i$ can be substituted by $u_i^H X u_i$, given that the generalized eigenvector is orthogonally weighted $u_i^H X u_i = 0 (i \neq k)$. Thus, (12) can be expressed in the form of (8), thereby proving Corollary 1.

*B. Theory of Linear Frequency-Modulation Signal Parameter Extraction*

1) **Theory for Extracting Initial Frequency and Frequency-Modulation Index of Linear Frequency-Modulated Signal**

**Definition 3.** Consider a linear frequency-modulated signal denoted by $y(t)$, which has the following form:

$$y(t) = a e^{j(kt^2 + 2\pi f t + \varphi_0)} \tag{13}$$

where $k, f, a, \varphi_0$ represent the corresponding modulation index, initial frequency, amplitude, and initial phase, respectively.

The first derivative of $y(t)$ is $\dot{y}(t)$. The Hankel matrices constructed from the discretized $y(t)$ and $\dot{y}(t)$ are denoted by $Y, \dot{Y}$ respectively, where

$$Y = \begin{pmatrix} y_1 & y_2 & \cdots & y_n \\ y_2 & y_3 & \cdots & y_{n+1} \\ \vdots & \vdots & \ddots & \vdots \\ y_n & y_{n+1} & \cdots & y_{2n-1} \end{pmatrix}, \dot{Y} = \begin{pmatrix} \dot{y}_1 & \dot{y}_2 & \cdots & \dot{y}_n \\ \dot{y}_2 & \dot{y}_3 & \cdots & \dot{y}_{n+1} \\ \vdots & \vdots & \ddots & \vdots \\ \dot{y}_n & \dot{y}_{n+1} & \cdots & \dot{y}_{2n-1} \end{pmatrix}, \tag{14}$$

Define $Y_H$ as follows:

$$Y_H = \begin{pmatrix} t_1 y_1 & t_2 y_2 & \cdots & t_n y_n \\ t_2 y_2 & t_3 y_3 & \cdots & t_{n+1} y_{n+1} \\ \vdots & \vdots & \ddots & \vdots \\ t_n y_n & t_{n+1} y_{n+1} & \cdots & t_{2n-1} y_{2n-1} \end{pmatrix}, \tag{15}$$

where $t_i$ is the temporal value that corresponds to element $y_i$ in Y.

Form (16) as follows[38],[39]:

$$\dot{Y} v - (\lambda_L Y_H + \mu_L Y) v = 0. \tag{16}$$

**Proposition 2.** The solution to the multi-parameter eigenvalue system (16) for linear frequency-modulated signals contains $n$ repeated eigenvalues $(\lambda, \mu)$, given by the following equation:

$$(\lambda_L, \mu_L) = (j2k, j2\pi f) \tag{17}$$

$$\begin{aligned} &\dot{Y} u - (\lambda_L Y_H + \mu_L Y) u \\ &= (j2k Y_H + j\omega Y) u - (\lambda_L Y_H + \mu_L Y) u \\ &= ((j2k - \lambda_L) U_y D_y - (j\omega - \mu_L) U_H D_H) V_{yH} u \\ &= \begin{bmatrix} (j2k - \lambda_L)(u_y)_1 (d_y)_1 - (j\omega - \mu_L)(u_H)_1 (d_H)_1 \\ \vdots \\ (j2k - \lambda_L)(u_y)_i (d_y)_i - (j\omega - \mu_L)(u_H)_i (d_H)_i \\ \vdots \\ (j2k - \lambda_L)(u_y)_n (d_y)_n - (j\omega - \mu_L)(u_H)_n (d_H)_n \end{bmatrix} V_{yH} u \end{aligned} \tag{18}$$

Then if and only if $(\lambda_L, \mu_L) = (j2k, j2\pi f)$ :

$$(j2k - \lambda_L)(u_y)_i (d_y)_i - (j\omega - \mu_L)(u_H)_i (d_H)_i = 0 \tag{19}$$

Therefore, if $(\lambda_L, \mu_L) = (j2k, j2\pi f)$, there exists a non-zero solution $u$ to the null space of (16). This indicates the existence of a non-zero eigenvector $u$ that corresponds to the eigenvalue (j2k, j2πf) that satisfies (16). Hence, to multi-parameter eigenvalue theory, it can be inferred that (j2k, j2πf) denotes the two-parameter eigenvalue pair of (16). Thus, the demonstration of Proposition 2 is complete. To determine the solution for $(\lambda_L, \mu_L)$, one can construct another two-parameter eigenvalue equation with a similar form to (16) and create a system of equations with (16). The solution of this system can then be obtained utilizing the theory of multi-parameter eigenvalues. For detailed proof, please refer to Appendix B

2) **Theory of Extracting Component Amplitude and Phase from Linear Frequency-Modulated Signals**

**Definition 4.** Let a linear frequency-modulated signal be denoted as $l(t)$, with the following form:

$$l(t) = e^{j(kt^2 + 2\pi f t)}. \tag{20}$$

where $k$ and $f$ denote the corresponding frequency-modulation index and start frequency, respectively.

The Hankel matrix constructed from discretized *l(t)* is denoted as $L$, where

$$L = \begin{pmatrix} l_1 & l_2 & \cdots & l_n \\ l_2 & l_3 & \cdots & l_{n+1} \\ \vdots & \vdots & \ddots & \vdots \\ l_n & l_{n+1} & \cdots & l_{2n-1} \end{pmatrix} \tag{21}$$

Construct a generalized eigenvalue equation in the form of (3):

$$Y v = \lambda_{Ap} L v \tag{22}$$

**Corollary 2.** Once the start frequency and frequency-modulation index of the chirp signal are determined via Proposition 2, the amplitude and initial phase can be calculated using (19).

$$\lambda_{Ap} = a e^{j\varphi_0} \tag{23}$$

**Proof.** From the definitions of $Y$ and $L$, it can be concluded that

$$y_i = a e^{j\varphi_0} l_i \tag{24}$$

Thus, for $Y$ and $L$, each row $y_{C_0 i}$ and $l_{C_0 i}$ has the following relationship:

$$\begin{bmatrix} (\lambda_{Ap} - a e^{j\varphi_0}) l_1 \\ \vdots \\ (\lambda_{Ap} - a e^{j\varphi_0}) l_i \\ \vdots \\ (\lambda_{Ap} - a e^{j\varphi_0}) l_n \end{bmatrix} v = 0 \tag{25}$$

That is, when $\lambda_{Ap} = a e^{j\varphi_0}$, there exists a non-zero vector that satisfies $Y v = \lambda_{Ap} L v$ . According to generalized eigenvalue theory, $a e^{j\varphi_0}$ is the generalized eigenvalue of $Y v = \lambda_{Ap} L v$. Moreover, the magnitude and phase angle of the generalized eigenvalue $\lambda_{Ap}$ correspond to the amplitude $a$ and initial phase $\varphi_0$ of the linear frequency-modulated signal, respectively, thus proving Corollary 2.

### *C. Algorithm Implementation*

1. SNGEM signal processing algorithm:

**Algorithm 1:** SNGEM signal processing algorithm

**Input:** Multi-frequency time-domain signal

**Output:** The frequency, amplitude, and initial phase of the components

1: Sampling the processed multi-frequency time-domain signal and its derivative signal.
2: Construct a generalized eigenvalue equation of the form $\Psi u = \lambda X u$ based on the theories discussed in Section II.A to calculate the frequencies of the components.
3: Estimate generalized eigenvalues of singular matrix pencil via SVD.
4: Calculate the frequency amplitudes and initial phases of the components according to the theories discussed in Section II.A.

2. SNGEM signal processing algorithm:

**Algorithm 2:** SNGEM signal processing algorithm:

**Input:** Linear frequency-modulation time-domain signal.

**Output:** Start frequency, FM parameters, amplitude, initial phase.

1: Preprocess the linear frequency-modulated (LFM) time-domain signal by filtering and noise reduction.
2: Sample the processed LFM signal and its analog derivative signal within the time interval $T_1$.
3: Similar to Step 2, but sampling is required over another time interval $T_2$.
4: Referring to the theories discussed in Section II.B, construct a system of equations formed by equations like $\dot{Y}v - (\lambda_L Y_H + \mu_L Y)v = 0$
5: Based on the theory of multi-parameter eigenvalues, solve the system of equations constructed in Step 4 to determine the starting frequency and frequency-modulation parameters.
6: Refer to Section II.B for the theory and solve the linear frequency-modulation amplitude and initial phase based on step 5.

## III. ALGORITHM VALIDATION

In the second section, we thoroughly explored and rigorously proved the theoretical foundation of the proposed algorithm for exactly extracting parameters under arbitrary sampling frequencies. Subsequent sections systematically investigate the performance of the algorithm by examining its statistical properties and evaluating its effectiveness in practical applications. The subsequent sections of this study are organized as follows:

First, we theoretically derive and simulate the feasibility and robustness of the proposed algorithm. The ADS tool suite verifies the feasibility of extracting the first-order derivative of the signal using a first-order Butterworth high-pass filter with a microstrip line and examines the effects of undersampling and noise on algorithm performance.

Although the ADS simulation results in the experimental part show that the microstrip line high-pass filter can effectively extract the first derivative of the signal in a wide frequency band, a commercial device in the initial stage of the experiment in the 2GHz Frequency range is used owing to the lack of precise microstrip line processing technology. Future research will focus on developing precision processing technology for SNGEM-related microstrip line filters.

Based on the obtained data, this study comprehensively compared the performance of the SNGEM method with that of IpFFT, CSS and SFT in terms of broadband spectrum sensing, and highlighted the advantages of SNGEM in spectrum sensing tasks. Additionally, this study investigated the application of the MEM algorithm in processing simulated GNSS Doppler signals of high-speed aircraft to extract speed and acceleration information, and verified the effectiveness and adaptability of the algorithm in solving dynamic signal extraction problems.

### *A. Algorithm feasibility analysis and verification.*

To assess the feasibility of implementing SNGEM on hardware platforms, we employed the ADS tool suite for preliminary research into acquiring the first-order differential of radio frequency signals using a first-order Butterworth high-pass filter. Fig. 2 illustrates the simulation, as well as the schematic for the corresponding microstrip line filter implementation. The results suggest that within a defined frequency range, the amplitude-frequency response of the first-order Butterworth high-pass filter exhibits linearity and dispersion that can be approximated as the ratio of the signal's first-order differential to the cutoff frequency. Furthermore, our findings alleviate concerns regarding abnormal phenomena that may arise from the amplification of high-frequency noise, as after the cutoff frequency, the filter's amplitude-frequency response tends towards unity, precluding such issues.

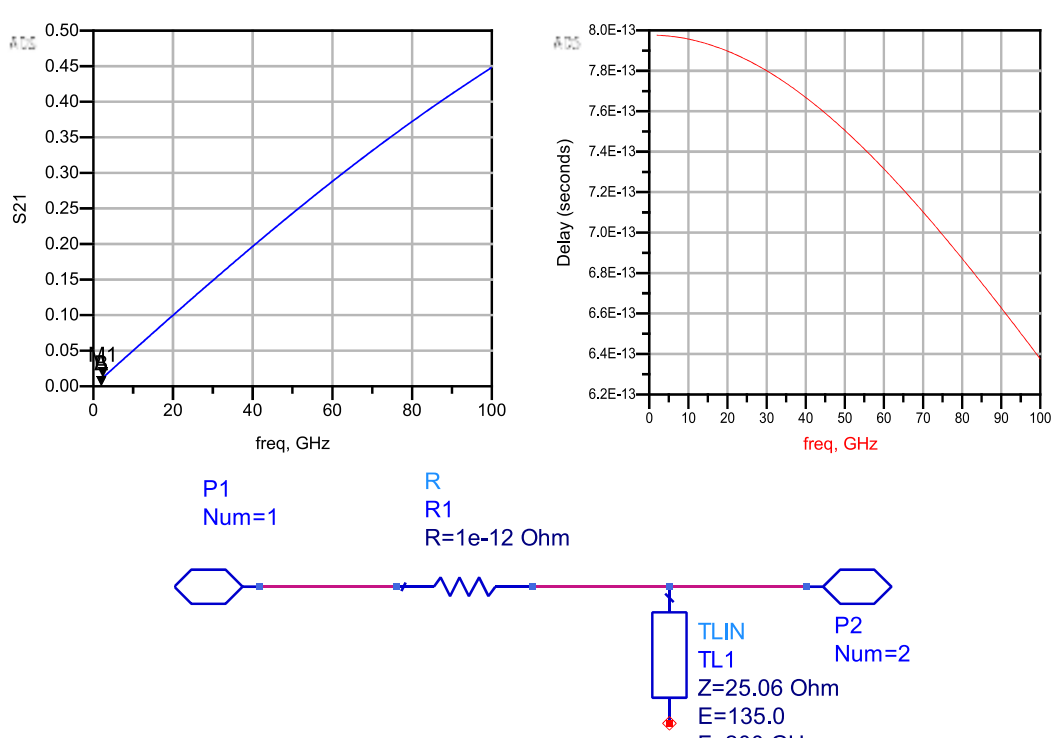

**Fig. 2** ADS Simulation of First-Order Differential Filter.

### B. Algorithmic robustness analysis

First, we discuss the matching of the number of sampling points to the order of the signal. When sampling sparse signals in band with existing hardware, it is not difficult to increase the number of sampling points, usually to more than twice the magnitude of the signal. In this case, the constructed matrix is singular when there is no noise. The SVD method can be used to effectively extract the corresponding generalized eigenvalues[32]. Details of the relevant experiments can be found in Sect. III.C and Sect. III.D. In some special cases, when the number of sampling points is less than twice the order of the signal, the Hankel matrix constructed from these sampling points can be considered a submatrix of the matrix satisfying the requirements. According to the Cauchy interpolation theorem for generalized eigenvalues, the actually calculated generalized eigenvalues should be interpolated from the real values. However, we can consider the signal as follows.

$$S_{truncated} = S_{main} + S_{noise} \tag{26}$$

Owing to the selective effect of SVD, we can always obtain an acceptable approximation of the main components, which is also of great importance in practical applications.

Finally, we examine the effect of noise on the generalized eigenvalues. Due to the orthogonality of the signal space and the noise space, this is not the case when examining the frequency of the signal outside the poles of the filter size frequency response, i.e. h. If the size of the signal components is not reduced excessively, it is difficult to separate the principal components from noise using singular value decomposition[32].

Fig. 3 illustrates the effectiveness of the proposed algorithm in extracting the signal power spectrum when the sample length is smaller than the signal order, including a simulation analysis performed in a noisy environment. When the sample length is insufficient, the present study compares the discrepancy between the principal component frequency that corresponds to the matrix order and the extraction frequency, including the overall divergence between the reconstructed signal and the original signal. Panel A shows that SNGEM can effectively extract the primary components of the power rate signal despite a limited sample length. In a noisy environment, the algorithm presented in this paper has high extraction accuracy at high SNR . However, at low SNR, the extraction accuracy of the algorithm needs to be improved, which could provide a potential avenue for future research.

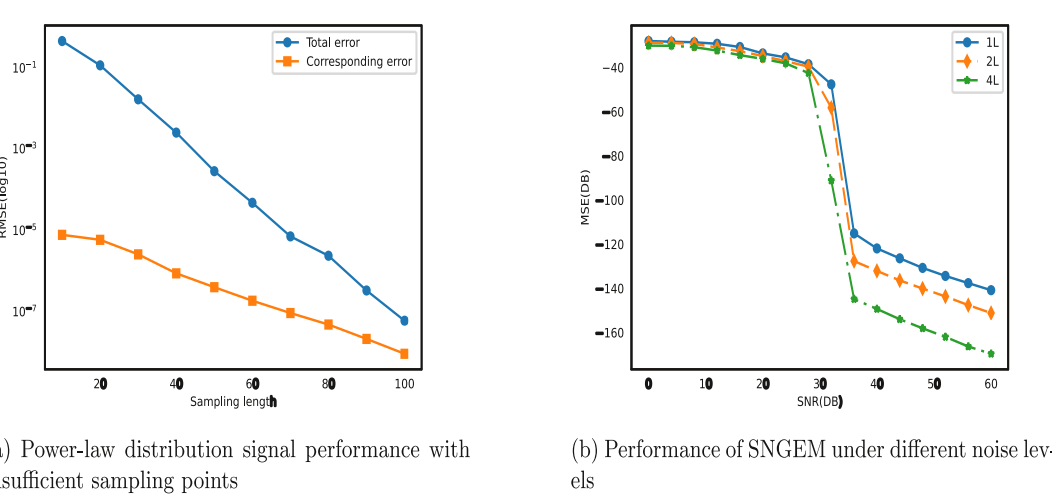

(a) Power-law distribution signal performance with insufficient sampling points

(b) Performance of SNGEM under different noise levels

**Fig. 3** Robustness Analysis of SNGEM Algorithm.

### C. Comparative Analysis of SNGEM and IpFFT Extraction Accuracy

In the absence of noise, we compared the accuracy of the parameters extracted using the proposed algorithm and the IpFFT method at various sampling frequencies. The number of components of the signal is $m$ = 10, and the true frequency $f$, amplitude $A$, and phase are presented in Table I.

TABLE I
THEORETICAL REFERENCE VALUES FOR MULTI-FREQUENCY SIGNAL COMPONENT PARAMETERS

| N | 1 | 2 | 3 | 4 | 5 | 6 | 7 | 8 | 9 | 10 |
|---|---|---|---|---|---|---|---|---|---|---|
| ***f*(Hz)** | 935 | 957 | 1297 | 1317.5 | 3120 | 3135 | 4460 | 5530 | 5970 | 7990 |
| ***A*** | 1.5 | 3.5 | 2 | 0.1 | 1.2 | 0.8 | 2.5 | 0.8 | 1 | 0.3 |
| $\Phi_i$ **(°)** | 30 | 50 | 170 | 230 | 90 | 145 | 360 | 330 | 280 | 360 |

The sampling frequencies are 2.5 times for Zoom-IpFFT and approximately 0.01 times the maximum component frequency for SNGEM. The SNGEM sample length is 38, which is 2m-1 of the original signal and the filtered signal according to Proposition 1. It is well known that spectral leakage and the picket fence effect degrade the spectral resolution of Zoom-IpFFT. To address this problem, scholars have proposed various improvement methods, such as the interpolated Fourier transform and Zoom-IpFFT. The interpolated Fourier transform enhances the frequency analysis accuracy by performing interpolation operations at the main lobe peak of the spectrum and its adjacent points. Zoom-IpFFT, on the other hand, improves the analysis accuracy by refining the local frequency domain of the signal. However, both methods inherently require a longer equivalent signal sampling length. Therefore, we adopted an equivalent method, which involves using signals longer than the SNGEM to compare the spectral resolutions of the two methods. The Zoom-IpFFT sample length is 1024. The results are presented in Table II.

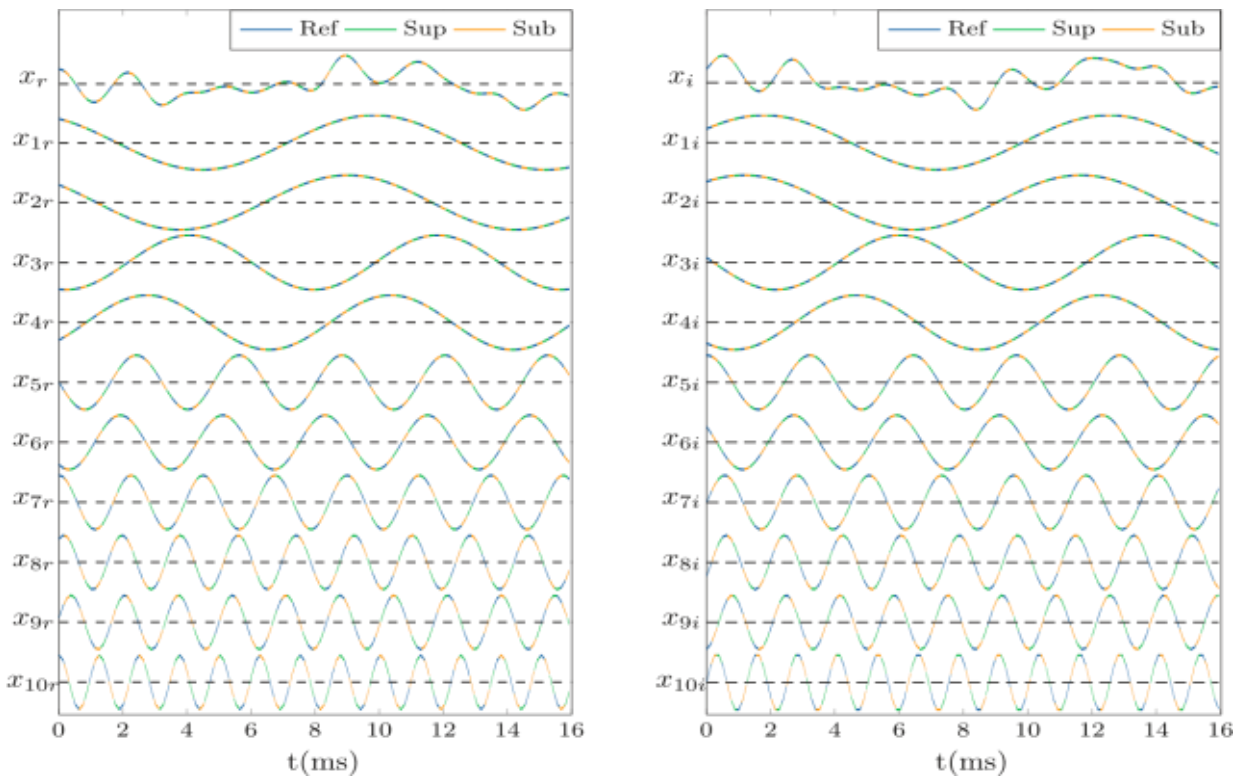

a: Real parts b:Imaginary parts

**Fig. 4** Multi-frequency complex signals and their component signals, where $x_r$ and $x_i$ correspond to the time-domain signals of the real and imaginary parts of the original signal.

TABLE II

COMPARISON OF ZOOM-IPFFT AND SNGEM FOR MULTI-FREQUENCY SIGNAL COMPONENT PARAMETERS

| N | Relative error of Zoom-IpFFT | | | Relative error of SNGEM methods with Nyquist sampling | | | Relative error of SNGEM methods with truly Sub-Nyquist sampling | | |
|---|---|---|---|---|---|---|---|---|---|
| | $\Delta f_i$ | $\Delta A_i$ | $\Delta \Phi_i$ | $\Delta f_i$ | $\Delta A_i$ | $\Delta \Phi_i$ | $\Delta f_i$ | $\Delta A_i$ | $\Delta \Phi_i$ |
| 1 | 1.22E-11 | 5.26E-06 | 1.12E-05 | 7.57E-12 | -4.64E-08 | 8.10E-09 | 5.66E-12 | -7.83E-07 | 5.44E-09 |
| 2 | 1.51E-11 | 7.61E-07 | 6.73E-06 | 2.10E-12 | 3.44E-08 | -4.07E-08 | 1.47E-12 | 7.35E-08 | -3.35E-09 |
| 3 | 2.18E-11 | 8.14E-06 | 1.11E-05 | 1.75E-12 | -1.09E-07 | -1.83E-09 | 1.62E-12 | -7.86E-08 | -8.31E-10 |
| 4 | 1.27E-11 | 2.22E-06 | 5.40E-06 | 5.79E-12 | -4.92E-09 | 1.14E-08 | 4.14E-12 | -6.85E-07 | 1.95E-08 |
| 5 | 4.78E-12 | 2.20E-06 | 7.87E-06 | 4.82E-12 | -1.01E-07 | 5.97E-10 | 1.53E-12 | -9.74E-08 | 9.04E-10 |
| 6 | 7.36E-12 | 9.12E-09 | 8.94E-06 | 1.99E-12 | -6.30E-09 | 1.36E-09 | 2.24E-12 | -2.05E-08 | 2.33E-09 |
| 7 | 3.81E-12 | 4.48E-07 | 2.22E-06 | 1.75E-12 | 2.67E-09 | -1.54E-10 | -1.32E-12 | 3.76E-07 | -3.30E-10 |
| 8 | 1.46E-11 | 2.24E-08 | 4.56E-06 | 8.39E-13 | -1.29E-08 | 1.66E-09 | 1.10E-12 | -1.53E-08 | 7.48E-10 |
| 9 | 1.08E-11 | 1.76E-06 | 8.15E-06 | -7.96E-13 | 5.06E-08 | -4.71E-07 | 6.24E-12 | 3.18E-08 | -2.19E-07 |
| 10 | 5.60E-12 | 3.15E-06 | 2.91E-07 | -1.58E-12 | -4.23E-08 | 1.00E-08 | -1.20E-12 | -4.67E-07 | 7.84E-09 |

Table II reveals that the largest errors primarily originate from adjacent signal components with the smallest frequency intervals and the greatest amplitude differences. These errors are attributed to both the algorithm and inherent limitations in numerical calculations performed by computers. By contrast, the errors for other components in the table are significantly smaller. Evidently, the proposed SNGEM method can provide more accurate results than the Zoom-IpFFT method when extracting the frequency, amplitude, and phase of signal components, even under sub-Nyquist sampling.

Fig. 4 presents a comparison between the signal components reconstructed from the parameters extracted by the SNGEM and true values. The legend "Ref" represents the theoretical values, while "Sup" and "Sub" correspond, respectively, to reconstruction results that satisfy the Nyquist sampling theorem and sub-Nyquist sampling conditions.

In Fig. 4, $x_r$ and $x_i$ represent the real and imaginary parts of the signal $x(t)$, respectively, whereas $x_{1r}$ to $x_{10r}$ denote the real parts of components 1 through 10, where $x_{1i}$ to $x_{10i}$ are their corresponding imaginary parts. It is evident from the Fig. that the computational results of the SNGEM are highly consistent with the theoretical values, regardless of whether the Nyquist sampling theorem is satisfied. This result visually confirms the accuracy and reliability of the SNGEM at various sampling frequencies, further validating its effectiveness for parameter extraction.

*D. Application Simulation*

1) **Application of SNGEM Method in Wideband Spectrum Sensing**

We compared the accuracy of frequency, amplitude, and phase extraction using the SNGEM with CSS and SFT to validate the performance of the proposed algorithm in extraction of wideband signals. In the experiment, the amplitude and phase parameters of the signal are consistent with those listed in Table I. The frequency component was amplified to 1e5 times its original value, with the highest frequency component being approximately 0.8 GHz. The original signal sample length $L$ for CSS and SFT ranged from 2048 to 16384, with a step size of 1024 and a sampling frequency of 2.1 times the highest component frequency. To ensure perfect reconstruction by CSS, both the compression ratio and measurement length must satisfy the principles of perfect reconstruction [40]. The bucket parameter B of SFT must be an integer approximately equal to $\sqrt{KL}$[41]. The length of the filtered signal was 6B, and the final compressed sample length was the maximum of CSS and SFT. For SNGEM, the sampling rate was 0.001 times the Nyquist sampling frequency, with sample lengths ranging from 76 to 608 points, a step size of 38, and 2000 simulations[42]. The simulation results are depicted in Fig.. 6.

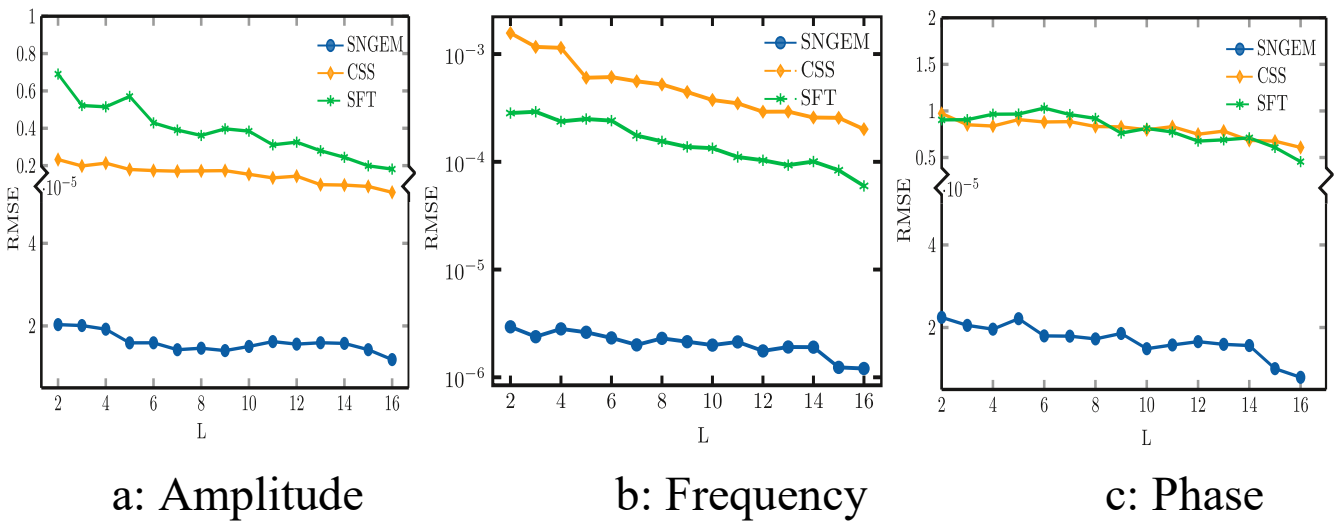

a: Amplitude b: Frequency c: Phase

**Fig. 5** Comparison of signal parameter extraction accuracy, where SNGEM, CSS, and SFT correspond to the algorithm proposed in this paper, the compressed spectrum sensing algorithm, and the sparse Fourier, respectively. The Fig. shows the difference in the extraction accuracy of the amplitude, frequency, and phase of the signal components at different lengths.

The sub-Fig.s (a), (b), and (c) in Fig.. 5 compare the accuracy of amplitude, frequency, and phase extraction by SNGEM, CSS, and SFT, respectively. The horizontal axis represents the relative sample length used in the experiment, and the vertical axis represents the logarithm of the RMSE of the parameter extraction. Moreover, owing to the superior performance of SNGEM in amplitude and phase extraction compared to the comparative methods, a discontinuous coordinate axis method was used to display the results for these parameters.

Fig. 5 illustrates that for CSS and SFT, increasing the original sample length noticeably improves their frequency extraction accuracy, achieving approximately one order of magnitude improvement under the experimental conditions. Contrastingly, enhancements in amplitude and phase extraction precision were minimal. However, the proposed algorithm achieves higher accuracy in frequency, amplitude, and phase extraction with significantly lower sampling requirements—approximately 0.17 times the average length and 0.04 times the original signal length—thus outperforming standard CSS and SFT methods.

Note that while SNGEM can theoretically achieve accurate parameter extraction at very low sub-Nyquist sampling frequencies, it often requires increasing the duration of the signal. In practical applications such as digital transmission or GNSS positioning, each signal frame has a limited length. Therefore, the sampling time cannot exceed the frame length, imposing constraints on how low the subsampling frequency can practically be set, which is not infinitely small.[43],[44],[45]

2) **Application of SNGEM in Motion Information Extraction of Hypersonic Vehicles**

Global navigation satellite system (GNSS) receivers are widely deployed in both civilian and military domains, particularly in high-dynamic applications such as precision guidance. There exists a growing need for GNSS receivers to exhibit high reliability in highly dynamic environments. An essential aspect of capturing high-dynamic GNSS data involves accurately extracting the Doppler shift and its rate of change in the carrier signals. Owing to the characteristics of signals during rapid movements and changes, it is feasible to simplify these signals into a linear frequency-modulation (LFM) signal model during the data capture process[46],[47].

Accurately extracting the Doppler frequency and its rate of change from GNSS signals of high-speed moving targets not only corrects the frequency bias of GNSS positioning signals but also enables precise real-time inference of target motion states. Currently, velocity measurement methods based on GNSS Doppler shift are primarily categorized into two categories: derived and raw Doppler shift methods. The derived Doppler shift method offers higher accuracy in measuring the velocity of static targets, while the raw Doppler shift method excels in accuracy for targets that undergo non-uniform motion. This study focuses on utilizing the raw Doppler shift method.

The simulation parameters were set as follows: The GNSS carrier frequency is 1.5 GHz, the initial velocity is 2 Mach, the acceleration is 20 G, with two sampling times of 500 µs each, totaling approximately 1 ms, which corresponds to the length of a single GNSS signal[48],[49]. This configuration effectively avoids errors caused by phase jumps in the clock. Given that the distance between the aircraft and the satellite typically exceeds 20,000 km—significantly greater than the flight distance of the aircraft during the sampling interval, as illustrated in Fig. 6—the motion speed can be projected onto the line that connects the aircraft and the satellite at an angle $\theta_i$. Table III presents the simulation results. As indicated in the table, the extraction accuracy of SNGEM is more closely aligned with the theoretical value than that of the Wigner–Hough transform (WHT). This is because the WHT method employs the FFT transform in the time-frequency transformation, and its accuracy is also influenced by factors such as spectral leakage and the pickle fence effect.

TABLE III
DOPPLER SHIFT EXTRACTION OF GNSS SIGNALS FOR HIGH-SPEED MOVING TARGETS BASED ON GEM AND WHT

| $\theta_i$ (°) | Velocity(m/s) | | | | | Acceleration(m/s$^2$) | | | | |
|---|---|---|---|---|---|---|---|---|---|---|
| | Ref | SNGEM | Error of SNGEM | WHT | Error of WHT | Ref | SNGEM | Error of SNGEM | WHT | Error of WHT |
| 0 | 6.81E+02 | 6.81E+02 | 7.48E-06 | 6.86E+02 | 8.15E-03 | 1.96E+02 | 1.96E+02 | 5.28E-05 | 1.96E+02 | 3.67E-04 |
| 110 | -2.33E+02 | -2.33E+02 | 6.25E-08 | -2.33E+02 | 2.52E-03 | -6.71E+01 | -6.71E+01 | 4.56E-06 | -6.70E+01 | 1.03E-03 |
| 230 | -6.70E+02 | -6.70E+02 | 7.53E-07 | -6.76E+02 | 9.14E-03 | -1.93E+02 | -1.93E+02 | 4.62E-06 | -1.93E+02 | 2.72E-03 |
| 320 | 1.18E+02 | 1.18E+02 | 5.82E-08 | 1.17E+02 | 6.41E-03 | 3.41E+01 | 3.41E+01 | 7.49E-05 | 3.41E+01 | 8.15E-04 |

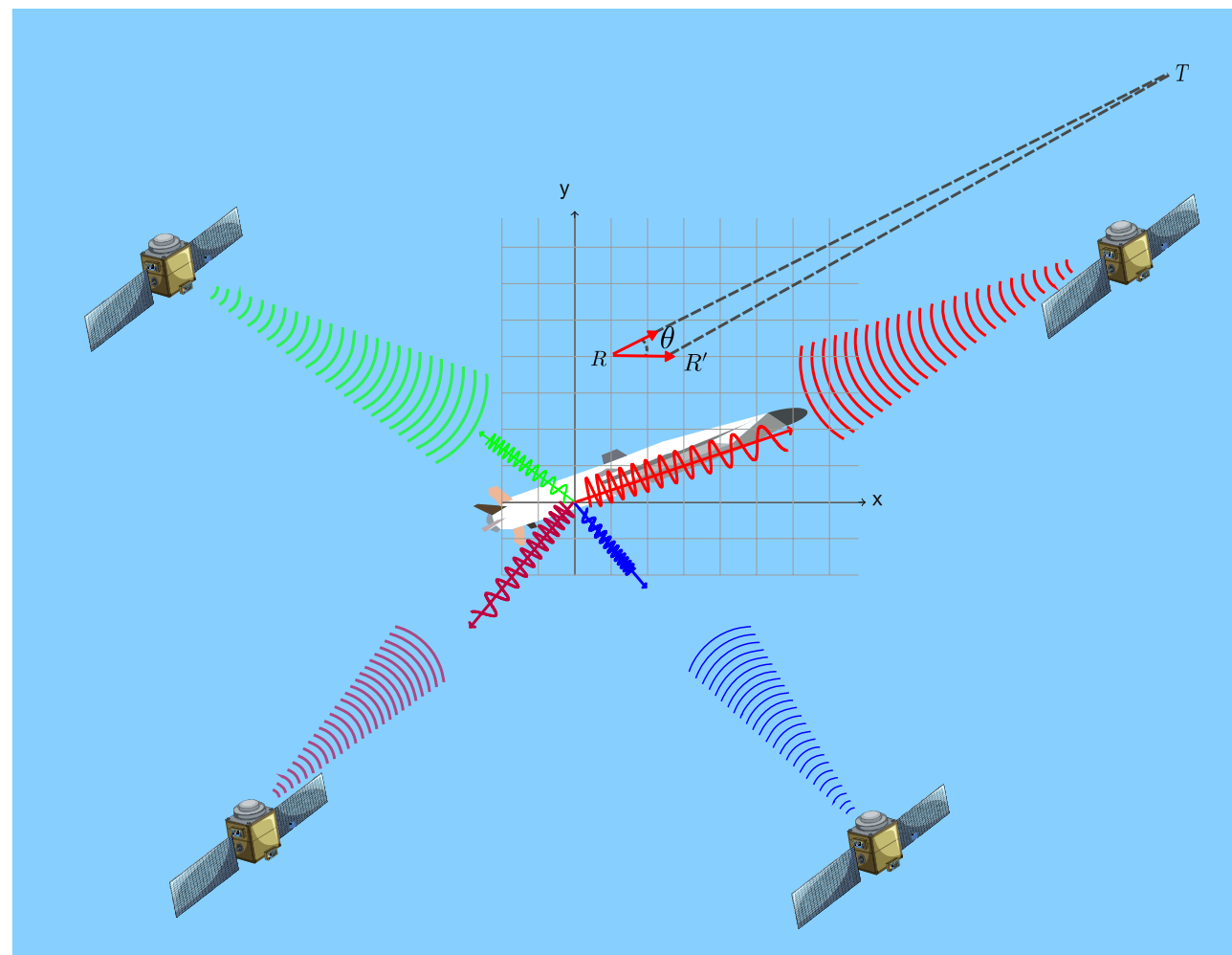

**Fig. 6** Doppler frequency offset parameter extraction for high-speed aircraft based on SNGEM, The Fig. shows the nonlinear Doppler shift of electromagnetic waves emitted by satellites located in different directions of the hypervelocity vehicle, which contains a wealth of information, including the velocity and acceleration of the vehicle.

## IV. Discussion

This paper proposed a novel generalized eigenvalue signal processing algorithm that enables super-resolution at extremely low sampling frequencies. Furthermore, we applied the theory of multi-parameter eigenvalues for the first time to accurately extract the parameters of LFM signals, providing rigorous theoretical proof of its viability. The subsequent discussion revolves around comparing algorithm performance, analyzing suitable applications, and examining hardware compatibility.

### *A. Mechanisms for Achieving High-Resolution with SNGEM Truly Sub-Nyquist Sampling*

Compared with CSS and SFT, the mechanism by which the SNGEM achieves super-resolution under truly sub-Nyquist sampling can be explained from different perspectives using theoretical frameworks based on generalized eigenvalue theory and compressed sensing.

Within the framework of generalized eigenvalue theory, the principle of the SNGEM is displayed in Fig. 7 in Appendix A: it leverages the non-coherence between rank-one matrices formed by multi-frequency signal components and their filtered signals. The framework employs a synchronous aliasing mechanism to effectively avoid the SL and picket-fence effect inherent in DFT, thus offering higher precision in super-resolution extraction. Moreover, SNGEM utilizes the density or infinite divisibility of its eigenvalues in the numerical domain to effectively circumvent the frequency collision problem present in the standard SFT.

Under the framework of compressed sensing, the principle of SNGEM operates as follows: The infinite sparse representation vectors of the original signal and its differentiated signal within the generalized multi-frequency framework maintain consistent non-zero element positions, changing only in their sparse coefficients. When solving for these coefficients, SNGEM does not directly determine the precise sparse coefficients of the original and differentiated signals. Instead, it computes the ratios of the sparse coefficients that share the same non-zero element positions, which correspond to the precise frequency values of the signal components. This method avoids problems such as destroyed sparsity due to SL and the picket-fence effect, which commonly affect other compressed sensing spectrum methods, thereby ensuring exact extraction of signal components.

It is crucial to emphasize that the necessary length of the measurement data for the SNGEM method is entirely determined by the number of signal components, reflecting the inherent sparsity of the signal. This characteristic enables SNGEM to accurately extract the parameters of signal components even with limited sample sizes, demonstrating its potential advantages in the realm of sub-Nyquist sampling techniques.

In practical applications such as cognitive radio systems and UWB communication, the SNGEM method offers significant advantages. In particular, it is well-suited when the signal bandwidth is wide and dynamic, standard sampling rates fail to satisfy the Nyquist criteria, and data collection time is limited. In many cases, signal frequencies are densely packed and dynamically changing, posing challenges for conventional methods that rely on extensive prior information. The SNGEM method excels by exactly extracting signal amplitudes and phases without heavy reliance on prior data. It efficiently retrieves multi-frequency signal components with high resolution from short data segments across varying sampling frequencies.

### *B. Hardware feasibility and compatibility*

The algorithm proposed in this study preprocesses the signal using a high-frequency low-pass filter or an UWB to reduce noise and prevent aliasing. Next, special filters are used to characterize the different frequency components of the signal to ensure that these components can be effectively distinguished. Many practical implementations of ultra-wideband filters[50],[51],[52], such as: Microstrip, coupled microstrip, coupled strip line and coplanar waveguide structures. Common filter designs include the $\lambda_0/4$ short-circuit line type with parallel prongs and the $\lambda_0/2$ open-circuit type with parallel prongs as well as their coupled line types[53]. Significant progress has also been made in the terahertz band in the research of waveguide filters, planar lightwave switching filters, MEMS filters, LTCC filters and photonic crystal filters[54],[55],[56].

While there are objections to detecting first-order differential signals, the application of differential sampling under low frequency conditions is widely accepted[57]. As shown in Section III.A, a first-order Butterworth high-pass filter coupled with a bandpass or microstrip filter can effectively capture first-order difference signals with a weighting coefficient within a specific frequency band. In addition, there is no infinite amplification of high frequency signals after the cutoff frequency. Compared to the advancement of specialized ultra-high frequency ADC devices and associated data transfer and back-end processing systems, the implementation of SNGEM hardware is significantly less complex. In addition, the hardware system proposed in this paper has remarkable scalability. In the case of a frequency band that is too wide, the signal can be divided into several frequency bands by using a bandpass filter and then the signal frequency can be converted into a frequency band within the operating range of existing devices by applying the mixing technique for filtering.

Compared to other random sampling methods in sub-Nyquist sampling techniques, the proposed algorithm utilizes uniform sampling for truly sub-Nyquist sampling. This approach facilitates a more practical and cost-effective hardware implementation solution. The required components include commercial-grade low-pass filters, first-order derivative samplers, standard low-rate sampling components, analog-to-digital converters (ADCs), and digital signal processors (DSPs). The implementation simplifies hardware design, reduces costs, and is straightforward to integrate into existing signal processing systems.

### *C. Issues and Concerns*

Importantly, although the SNGEM method provides a very low sampling rate, this is achieved at the expense of a longer sampling time. In certain applications, such as digital wireless communications, this may cause the sampling time to exceed the length of the unit data frame, which may result in skipping transitions requiring countermeasures.. Additionally, compared to traditional DFT-based CSS, our proposed method may have some limitations in terms of signal types that can be applied to continuous band signals. This one of the topics of our further research.

## V. CONCLUSION

This study proposed an innovative signal processing technique called the SNGEM, which operates at sub-Nyquist sampling frequencies 0.001 times the signal bandwidth and uses only 10% of the sampling length required by other compressed spectral sensing methods. SNGEM greatly improves reconstruction precision by approximately three to four orders of magnitude. The algorithm successfully achieved super-resolution reconstruction of multi-frequency complex signals at ultra-low sampling rates, and we fully validated its theoretical rationale. Furthermore, the SNGEM has been extended to extract characteristic parameters from LFM signals. From a hardware perspective, the proposed algorithm can be easily implemented using existing commercial devices without requiring new hardware development, thus significantly simplifying hardware design auto-comp and reducing costs. This opens up new opportunities for practical and commercial applications of compressed sensing technology.

Notably, while the SNGEM has clear advantages over Fourier-based CSS algorithms in accurately estimating characteristic parameters under sub-Nyquist sampling for standard sparse signals, and has successfully achieved precise extraction of LFM signals, its applicability across various signal types requires further research. Moreover, the robustness of the proposed algorithm under noisy conditions needs to be thoroughly assessed, which is crucial for ensuring its reliability in practical applications.

## APPENDIX A

**Proof.** For simplicity, (2.1) is simplified to the following form:

$$x(t) = \sum_{i=1}^{m} a_i \, e^{j(\omega_i t + \varphi_{i0})} \qquad (A.1)$$

Where $\omega_i = 2\pi f_i$

The filtered signal of $x(t)$ is:

$$\psi(t) = \sum_{i=1}^{m} A_i(f_i) \exp\left(j\Phi(f_i)\right) a_i e^{j(\omega_i t + \varphi_{i0})}. \qquad (A.2)$$

For the sake of simplicity, the following provisions are made:

$$s_i = e^{j\omega_i \Delta t}, \alpha_i = a_i e^{j\varphi_{i0}}, \beta_i = A(f_i)\exp\left(j\Phi(f_i)\right) \qquad (A.3)$$

By employing the definitions of $X, D_x$, and (A.1), (A.2), and (A.3) , we derive the following results:

$$X = \begin{pmatrix} \sum_{i=1}^{m} \alpha_i \, s_i & \sum_{i=1}^{m} \alpha_i \, s_i^2 & \cdots & \sum_{i=1}^{m} \alpha_i \, s_i^n \\ \sum_{i=1}^{m} \alpha_i \, s_i^2 & \sum_{i=1}^{m} \alpha_i \, s_i^3 & \cdots & \sum_{i=1}^{m} \alpha_i \, s_i^{n+1} \\ \vdots & \vdots & \ddots & \vdots \\ \sum_{i=1}^{m} \alpha_i \, s_i^n & \sum_{i=1}^{m} \alpha_i \, s_i^{n+1} & \cdots & \sum_{i=1}^{m} \alpha_i \, s_i^{2n-1} \end{pmatrix}$$

$$\Psi = \begin{pmatrix} \sum_{i=1}^{m} \beta_i \alpha_i s_i & \sum_{i=1}^{m} \beta_i \alpha_i s_i^2 & \cdots & \sum_{i=1}^{m} \beta_i \alpha_i s_i^n \\ \sum_{i=1}^{m} \beta_i \alpha_i s_i^2 & \sum_{i=1}^{m} \beta_i \alpha_i s_i^3 & \cdots & \sum_{i=1}^{m} \beta_i \alpha_i s_i^{n+1} \\ \vdots & \vdots & \ddots & \vdots \\ \sum_{i=1}^{m} \beta_i \alpha_i s_i^3 & \sum_{i=1}^{m} \beta_i \alpha_i s_i^{n+1} & \cdots & \sum_{i=1}^{m} \beta_i \alpha_i s_i^{2n-1} \end{pmatrix}. \qquad (A.4)$$

Subsequently, we investigate the diagonalization of $X, \Psi$ with Vandermonde decomposition[58][59]. For Hankel matrices $X_i$ on the components of $x(t)$,

$$X_i = \begin{pmatrix} \alpha_i s_i & \alpha_i s_i^2 & \cdots & \alpha_i s_i^n \\ \alpha_i s_i^2 & \alpha_i s_i^3 & \cdots & \alpha_i s_i^{n+1} \\ \vdots & \vdots & \ddots & \vdots \\ \alpha_i s_i^n & \alpha_i s_i^{n+1} & \cdots & \alpha_i s_i^{2n-1} \end{pmatrix} = (1, s_i, \cdots, s_i^{n-1})^T \alpha_i s_i (1, s_i, \cdots, s_i^{n-1}) \quad (A.5)$$

where

$$S_{ci} = (1 \quad s_i \quad \cdots \quad s_i^{n-1}). \quad (A.6)$$

The matrix $X$ can be represented as follows:

$$X = \sum_{i=1}^{m} X_i = \sum_{i=1}^{m} S_{ci}^T \begin{pmatrix} \alpha_1 s_1 & & & \\ & \alpha_2 s_2 & & \\ & & \ddots & \\ & & & \alpha_m s_m \end{pmatrix} S_{ci} = S_c^T \begin{pmatrix} \alpha_1 s_1 & & & \\ & \alpha_2 s_2 & & \\ & & \ddots & \\ & & & \alpha_m s_m \end{pmatrix} S_c \quad (A.7)$$

and

$$S_c = [S_{c1}^T, S_{c2}^T, \cdots, S_{cm}^T]^T, \quad (A.8)$$

where $S_c$ represents the Vandermonde matrix of the Hankel matrix $X$ formed by the discrete multiple-frequency complex signal.

Performing a Vandermonde decomposition of the Hankel matrix $\Psi$ with discrete components $\psi(t)$:

Thus, from (A.6) and (A.8), $\Psi$ can be expressed as follows:

$$\Psi = \sum_{i=1}^{m} \Psi_i = S_c^T \begin{pmatrix} \beta_1 \alpha_1 s_1 & & & \\ & \beta_2 \alpha_2 s_2 & & \\ & & \ddots & \\ & & & \beta_m \alpha_m s_m \end{pmatrix} S_c \quad (A.9)$$

$$\Psi v - \lambda X u = S_c^T \begin{pmatrix} (\lambda - \beta_1) s_1 & & & \\ & (\lambda - \beta_2) s_2 & & \\ & & \ddots & \\ & & & (\lambda - \beta_m) s_m \end{pmatrix} S_c u \quad (A.10)$$

Equation (A.10) shows that the rank of $\Psi - \lambda X$ decreases from $m$ to $m$–1 only when $\lambda = \beta_i$. This implies the existence of a vector $v$ corresponding to $\beta_i$, which satisfies the condition for a non-zero solution to $(\Psi - \lambda X)u = 0$. According to the definition of generalized eigenvalues, $\beta_i$ represents the generalized eigenvalue of (3). Therefore, the proof of Proposition 1 is complete.

Then for $\beta_i$, given a specific filter, since its amplitude-frequency characteristic is always known, we can extract the corresponding frequency $f_i$ by performing the inverse transformation of the amplitude $A(f_i)$ of the generalized eigenvalue $\beta_i$.

The mechanism of the SNGEM can be explained as shown in Fig. 7:

- At first, as illustrated in the above Fig.7, the signal designated as $x(t)$ is the one to be decomposed. The dark blue curve represents the original signal, and the sinusoidal curves of varying colors represent the individual signal components. The signal $x(t)$ is then filtered in the time domain to obtain the filtered signal $\psi(t)$. Based on the properties of the filtered operation, the frequencies of the filtered signal $\psi(t)$ and the components of the original signal $x(t)$ remain consistent.
- Subsequently, as illustrated in the upper portion of the middle section, following the sampling or discretization of $x(t)$, the corresponding Hankel matrix $X$ is constructed. The matrices of varying colors represent distinct components. The Hankel matrix $\Psi$, constructed from the difference signal

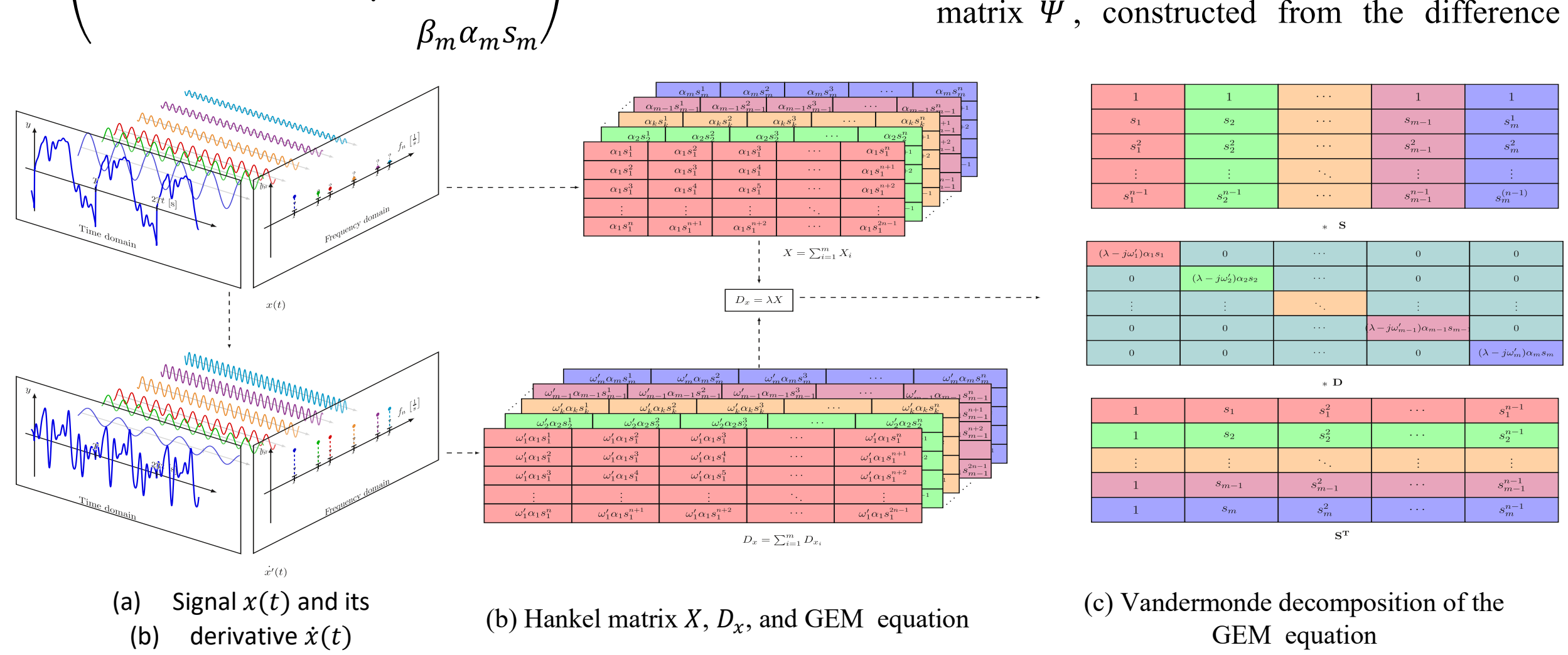

(a) Signal $x(t)$ and its (b) derivative $\dot{x}(t)$ (b) Hankel matrix $X$, $D_x$, and GEM equation (c) Vandermonde decomposition of the GEM equation

**Fig. 7** Mechanism of GEM, mechanism of the generalized eigenvalue method, and the meanings and explanations of each part are detailed in the main section.

Substituting (A.6) and (A.9) into (3) yields:

$\psi(t)$, is shown in the lower portion of the middle section.

- Finally, as illustrated in the Fig. on the right, the

Hankel matrix on both sides of the generalized eigenvalue equation is decomposed using Vandermonde decomposition. Subsequently, the corresponding terms are combined to transform the generalized eigenvalue equation into the product of a Vandermonde matrix and a diagonal matrix. The Vandermonde matrix is represented by $S$, and its transpose is denoted by $S^T$. $\Psi$ represents a diagonal matrix that contains the generalized eigenvalues and the angular frequency information of the components.

Appendix B

**Proof.** It follows from (9) that

$$\dot{y}(t) = j(2\mathrm{kt} + 2\pi f)ae^{j(kt^2+2\pi ft+\varphi_0)} \qquad (B.1)$$

For simplicity, let $\omega = 2\pi f$, then by the definition of y and (B.1):

$$\dot{Y} = \begin{pmatrix} \dot{y}_1 & \dot{y}_2 & \cdots & \dot{y}_n \\ \dot{y}_2 & \dot{y}_3 & \cdots & \dot{y}_{n+1} \\ \vdots & \vdots & \ddots & \vdots \\ \dot{y}_n & \dot{y}_{n+1} & \cdots & \dot{y}_{2n-1} \end{pmatrix}$$

$$= j\begin{pmatrix} (2kt_1+\omega)y_1 & (2kt_2+\omega)y_2 & \cdots & (2kt_n+\omega)y_n \\ (2kt_2+\omega)y_2 & (2kt_3+\omega)y_3 & \cdots & (2kt_{n+1}+\omega)y_{n+1} \\ \vdots & \vdots & \ddots & \vdots \\ (2kt_n+\omega)y_n & (2kt_{n+1}+\omega)y_{n+1} & \cdots & (2kt_{2n-1}+\omega)y_{2n-1} \end{pmatrix} \qquad (B.2)$$

$$= j2kY_H + j\omega Y.$$

We perform right-joint SVD on matrices $Y$ and $Y_H$,[60][61] which yields

$$Y = U_y D_y V_{yH}, Y_H = U_H D_H V_{yH}. \qquad (B.3)$$

Here, the left singular vector matrices for $y$ and $y_H$ in the right-joint SVD are denoted as $U_y$ and $U_H$, respectively. The associated singular vectors are represented as $(u_y)_i, (u_H)_i$. Diagonal matrices $D_y, D_H$ represent the diagonal elements in the right-joint SVD. The diagonal elements are denoted as $(d_y)_i, (d_H)_i$. Matrix $V_{yH}$ represents the right singular vectors that are common in the right-joint SVD.

Substituting (B.2) and (B.3) into (12) yields

$$\dot{Y}u - (\lambda_L Y_H + \mu_L Y)u$$
$$= (j2kY_H + j\omega Y)u - (\lambda_L Y_H + \mu_L Y)u$$
$$= ((j2k - \lambda_L)U_y D_y - (j\omega - \mu_L)U_H D_H)V_{yH}u \qquad (B.4)$$

$$= \begin{bmatrix} (j2k-\lambda_L)(u_y)_1(d_y)_1 - (j\omega-\mu_L)(u_H)_1(d_H)_1 \\ \vdots \\ (j2k-\lambda_L)(u_y)_i(d_y)_i - (j\omega-\mu_L)(u_H)_i(d_H)_i \\ \vdots \\ (j2k-\lambda_L)(u_y)_n(d_y)_n - (j\omega-\mu_L)(u_H)_n(d_H)_n \end{bmatrix} V_{yH}u$$

Then if and only if $(\lambda_L, \mu_L) = (j2k, j2\pi f)$ :

$$(j2k - \lambda_L)(u_y)_i(d_y)_i - (j\omega - \mu_L)(u_H)_i(d_H)_i = 0 \quad (B.5)$$

Therefore, if $(\lambda_L, \mu_L) = (j2k, j2\pi f)$, there exists a non-zero solution $u$ to the null space of (B.4). This indicates the existence of a non-zero eigenvector $u$ that corresponds to the eigenvalue (j2k, j2πf) that satisfies (12). Hence, to multi-parameter eigenvalue theory, it can be inferred that (j2k, j2πf) denotes the two-parameter eigenvalue pair of (12). Thus, the demonstration of Proposition 2 is complete. To determine the solution for $(\lambda_L, \mu_L)$, one can construct another two-parameter eigenvalue equation with a similar form to (12) and create a system of equations with (12). The solution of this system can then be obtained utilizing the theory of multi-parameter eigenvalues.

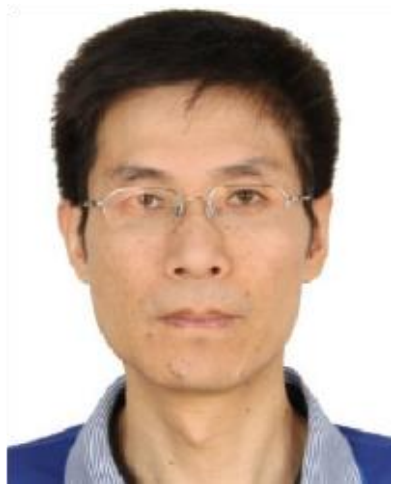

**Baoguo Liu** received the B.S. degree in grain machinery from Zhengzhou Grain University, Zhengzhou, China, in 1982, the M.S. degree in solid mechanics from the Dalian University of Technology, Dalian, China, in 1986, and the Ph.D. degree in solid mechanic from Chongqing University, Chongqing, China, in 2002. Since 2004, he has been a Professor with the Henan University of Technology. He has authored three books and more than 50 articles. His research interests include signal processing, fault diagnosis and motor dynamics. He has presided four NSFC projects and more than ten other scientific research projects. He was China’s National 863 Project, the National Natural Science Foundation of China, China’s National Youth Thousand Assessment Experts.

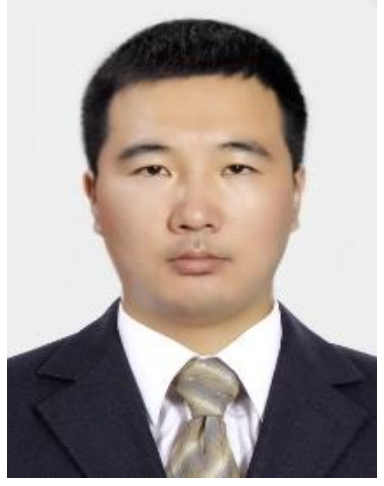

**Huiguang Zhang** graduated from Southwest Jiaotong University (SWJTU) in 2007 with a Bachelor's degree in Engineering and graduated from SWJTU in 2010 with a Master's degree. He is currently pursuing his Ph.D. degree at Henan University of Technology. His research interests include signal processing.

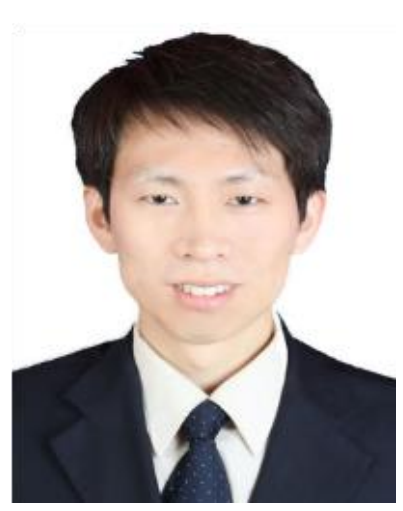

**Wei Feng** received the B.S. degree in mechanical engineering and automation from the Wuhan University of Technology, Wuhan, China, in 2004, the M.S. degree in mechanical engineering from Xi'an Jiaotong University, Xian, China, in 2011, the Ph.D. degree in machine building and automation from Xiamen University, Xiamen, China, in 2016. Since 2016, he has been an Assistant Professor of mechanical engineering with the Henan University of Technology. He is the author of 25 articles. His research interests include signal processing and mechanical dynamics.